\documentclass[twocolumn,showpacs,preprintnumbers]{revtex4}
\usepackage{amssymb}
\usepackage{amsmath}
\usepackage{graphicx}
\usepackage{dcolumn}
\usepackage{bm}
\usepackage{color}
\usepackage{soul}

\soulregister\cite7
\soulregister\ref7

\begin{document}

\title{Multiple ground-state instabilities in the anisotropic quantum Rabi model}
\author{Xiang-You Chen$^{1}$, Liwei Duan~$^{1,2}$, Daniel Braak~$^{3,\dagger}$, and Qing-Hu Chen~$^{1,4, *}$}

\affiliation{
$^{1}$ Zhejiang Province Key Laboratory of Quantum Technology and Device, Department of Physics, Zhejiang University, Hangzhou 310027, China \\
$^{2}$ Department of Physics, Zhejiang Normal University, Jinhua 321004,  China \\
$^3$ EP VI and Center for Electronic Correlations and Magnetism, University of Augsburg, 86135 Augsburg, Germany \\
$^4$   Collaborative Innovation Center of Advanced Microstructures, Nanjing University, Nanjing 210093, China
 }\date{\today}

\begin{abstract}
In this work, the anisotropic variant of the quantum Rabi model with different coupling
strengths of the rotating and counter-rotating wave terms is studied by the
Bogoliubov operator approach. The anisotropy preserves the parity symmetry of the original model. We derive the corresponding  $G$-function, which yields both
the regular and exceptional eigenvalues. The exceptional eigenvalues
correspond to the crossing points of two energy levels with
different parities and are doubly degenerate. We find analytically that the ground-state and the first excited state can cross
several times, indicating multiple first-order phase transitions as function of the coupling strength. These crossing points are related to manifest parity symmetry of the Hamiltonian, in contrast to the level crossings in the asymmetric quantum Rabi model which are caused by a hidden symmetry.
\end{abstract}

\maketitle

\section{Introduction}

The Quantum Rabi model (QRM) describes a two-level system coupled to a
single electromagnetic mode (an oscillator) via the dipole term~\cite{Rabi},
the simplest form of light-matter interaction. As such it has many
applications in numerous fields ranging from quantum optics and quantum
information science to condensed matter physics. In conventional (weak
coupling) applications to quantum optics, the rotating-wave (RW) terms are
kept and the counter-rotating-wave (CRW) terms are neglected~\cite{Scully}.
This so-called rotating-wave approximation (RWA) is equivalent to the
Jaynes-Cummings model~\cite{Jaynes-Cummings}, which is solvable in closed
form.

Over the past decade, developments in circuit QED~\cite{Wallraff,Deppe} have
allowed to reach the ultra-strong coupling regime where the coupling between
the superconducting qubit and the resonator can reach $10\%$ of the mode
frequency $\omega$. In this ultrastrong-coupling regime, evidence for the
breakdown of the RWA and the importance of CRW terms has been provided by
measurements of transmission spectra~\cite{Niemczyk,exp}.
More recently, even the deep strong coupling region has been realized
experimentally, where the coupling strength is of the same order as the mode
frequency~\cite{Yoshihara}. In this regime, the RWA cannot even
qualitatively describe the system~\cite{Casanova}. Although the spectrum of
the full QRM is very easy to obtain numerically by working in a truncated
bosonic Hilbert space, the exact analytical solution and with it qualitative
statements about the spectrum are difficult to obtain compared to the
super-integrable Jaynes-Cummings model. Analytical approximations have been
obtained at different levels, such as the limit of small energy splitting $%
\Delta$ of the qubit ($\Delta /\omega <0.5$) and the deep strong coupling
regime~ $g/\omega\sim 1$~\cite{Casanova,Hausinger,Ashhab}, weak and
intermediate coupling ($g/\omega <0.4$)~\cite{He}, and also in the whole
parameter range~\cite{Feranchuk,Irish,zhengh}. All these approximations,
while numerically often satisfying, miss some qualitative features of the
exact spectral graph like true level crossings and narrow avoided crossings.

An analytical solution based on the $\mathbb{Z}_2$-symmetry of the model and
using the Bargmann representation of $L^2(\mathbb{R})$ introduced a
transcendental function, called $G$-function in \cite{Braak}, whose zeros
yield the exact spectrum of the QRM. Shortly afterwards, it was found that
the $G$-function can be written in terms of Heun functions, known from the
theory of linear differential equations in the complex domain~\cite{Zhong}.
The $G$-function has a characteristic pole structure, giving information
about the form of the eigenstates and the distribution of the eigenvalues
along the real axis~\cite{Duan18,Braak19}. With its help, one may classify
the eigenvalues as belonging either to the regular or to the exceptional
spectrum, the former always non-degenerate, while the latter is comprised of
a degnerate and a non-degenerate part~\cite{Braak, Wakayama}. The $G$%
-function can be derived also in the more familiar Hilbert space $L^2(%
\mathbb{R})$, using the Bogoliubov operator approach, and thus in a
physically more intuitive way~\cite{Chen2012}.

The ``anisotropic'' generalization of the QRM where RW and CRW terms have
different coupling strengths (AiQRM) has been studied for quite a long time,
initially out of pure theoretical interest. It appeared first in the form of
the anisotropic variant of the Dicke model~\cite{Furuya}. Recently,
Goldstone and Higgs modes have been experimentally demonstrated in optical
systems with only a few (artificial) atoms, which can be described by the
anisotropic Dicke model with a small number of qubits~\cite{yejw2013}. This
experimental progress motivated theoretical studies of the anisotropic QRM
(a single qubit)~\cite{Fanheng,Tomka}. The AiQRM can also model a
two-dimensional electron gas with Rashba ($\alpha _R,$ corresponding to RW
coupling) and Dresselhaus ($\alpha _D,$ corresponding to CRW coupling)
spin-orbit interactions, subject to a perpendicular magnetic field~\cite%
{Erlingsson}. The two types of couplings can be tuned by external electric
and magnetic fields, allowing the exploration of the whole parameter space
of the model. It can also be directly realized in both cavity QED~\cite%
{Schiroa} and circuit QED~\cite{Wallraff}. For example, Ref.~\cite{Grimsmo}
proposes a realization of the AiQRM based on resonant Raman transitions in
an atom interacting with a high finesse optical cavity mode. { Very recently, it has been proposed that the  AiQRM can also be realized in the dispersive regime via momentum states instead of electronic states~\cite{Mivehvara}.}

The exact solution of the AiQRM has been obtained using the Bargmann
representation~\cite{Fanheng,Tomka}. The $G$-function was obtained by Xie
\textsl{et al}.~\cite{Fanheng}, and both regular and exceptional eigenvalues
have been studied. The isolated exact solutions at the level crossings (i.e.
a part of the exceptional spectrum) were found by Tomka \textsl{et al}.~\cite%
{Tomka}. The surprising finding of Ref.~\cite{Fanheng} was that for certain
parameter values the first excited state may form a degenerate doublet with
the ground state and belongs thus to the exceptional spectrum, which can
never happen in the isotropic QRM~\cite{braak-fmi}.

At this level crossing, the parity of the ground state changes sign and the
system undergoes a first order quantum phase transition~\cite{Fanheng}. In
the framework of the Bogoliubov operator approach, the anisotropic QRM has
been solved by two of the present authors \cite{Duan2015}. The doubly
degenerate exceptional solutions are explicitly given in Eq. (14) of Ref.
\cite{Duan2015} as a methodical alternative to Ref.~\cite{Tomka}, where the
problem was treated by a Bethe ansatz of Gaudin-Richardson type. Recently,
it has been claimed that the quantum phase transition of the AiQRM is
accompanied by the breaking of a hidden symmetry \cite{yingzj}.

The AiQRM continues to be an interesting topic, because it connects
continuously the Jaynes-Cummings model with the isotropic QRM. We revisit
the AiQRM along the lines of \cite{Duan2015} and focus on the crossings of
the first two energy levels in the spectral graph as function of increasing
coupling strength.

The paper is organized as follows. In Sec.~II, we construct the $G$-function
using extended coherent states. In Sec.~III, we analyze the exceptional
spectrum, which contains all possible level degeneracies, with the help of
the $G$-function.  We discuss the
crossings between the two lowest levels in Sec.~IV, and we obtain their
properties analytically. In this way, the ground state phase diagram of the
AiQRM in the plane spanned by the coupling strength and the anisotropy
parameter is obtained. We summarize our findings in Sec.~V.

\section{Model and analytic solutions}

The Hamiltonian of the anisotropic QRM reads~\cite{Fanheng,Tomka}
\begin{equation}
H=\frac{1}{2}\Delta \sigma _{z}+\omega a^{\dagger }a+g_{1}\left( a^{\dagger
}\sigma _{-}+a\sigma _{+}\right) +g_{2}\left( a^{\dagger }\sigma
_{+}+a\sigma _{-}\right) ,  \label{Hamiltonian}
\end{equation}%
where $\Delta $ is qubit level splitting, $a^{\dagger }$ $\left( a\right) $
is the photonic creation (annihilation) operator of the single radiation
mode with frequency $\omega$ (set to $1=\hbar=\omega$ in the following  and the figures), $g_{1}\ $%
and $g_{2}\ $ are the RW and CRW coupling constants respectively, and $%
\sigma _{k}(k=x,y,z)$ $\ $ are the Pauli matrices. Set $r=g_{2}/g_{1}\ $as
the anisotropic parameter and $g=g_{1}$ below.

The anisotropic QRM possesses the same $\mathbb{Z}_{2}$ symmetry as the
isotropic one. The parity operator is defined as $\hat{\Pi} =\exp \left(
i\pi \hat{N}\right)$, where $\hat{N}=a^{\dag }a+\sigma _{+}\sigma _{-}$ with
$\sigma _{\pm }=\left( \sigma _{x}\pm i\sigma _{y}\right) /2$ is the
operator of the total excitation number. Note that while $\hat{N}$ is not
conserved, the Hamiltonian \eqref{Hamiltonian} conserves its parity, ($\hat{N%
}\ \text{mod}\ 2$) and therefore $\hat{\Pi}$. The parity operator $\hat{\Pi}$
has two eigenvalues \ $\Pi =\pm 1$, corresponding to even and odd parity of $%
\hat{N}$.

We proceed to derive the $G$-function of \eqref{Hamiltonian}. Employing the
following transformation
\begin{equation}
P=\frac{1}{\sqrt{2}}\left(
\begin{array}{ll}
\sqrt{r} & ~1 \\
-\sqrt{r} & \;1%
\end{array}%
\right) ,\;\;P^{-1}=\frac{1}{\sqrt{2}}\left(
\begin{array}{ll}
\;\frac{1}{\sqrt{r}}\; & ~-\frac{1}{\sqrt{r}} \\
\;\;\;\;1 & \;\;\;\;\;1%
\end{array}%
\right) ,  \label{P1}
\end{equation}%
the Hamiltonian becomes
\begin{widetext}
\begin{equation}
H_{1}=PHP^{-1}=\left(
\begin{array}{ll}
a^{\dagger }a+\beta \left( a+a^{\dagger }\right) +\left( \frac{\lambda _{+}}{%
\beta }-\beta \right) a^{\dagger } & \;\;\;\;-\frac{1}{2}\Delta -\frac{%
\lambda _{-}}{\beta }a^{\dagger } \\
\;\;\;\;-\frac{1}{2}\Delta +\frac{\lambda _{-}}{\beta }a^{\dagger } &
\;a^{\dagger }a-\beta \left( a+a^{\dagger }\right) -\left( \frac{\lambda _{+}%
}{\beta }-\beta \right) a^{\dagger }%
\end{array}%
\right) ,
\end{equation}
\end{widetext}
where $\lambda _{\pm }=g^{2}\left( 1\pm r^{2}\right)/2$ and$\;\beta =g\sqrt{r%
}$.

We introduce two displaced bosonic operators with opposite displacements
\begin{equation}
A_{+}^{\dagger }=a^{\dagger }+\beta ;\;\;A_{-}^{\dagger }=a^{\dagger }-\beta
.  \label{dis}
\end{equation}
The bosonic number state in terms of the new photonic operators $%
A_{+}^{\dagger }$ and $A_{-}^{\dagger }$ are
\begin{eqnarray*}
\left| n\right\rangle _{A_{+}} &=&\frac{\left( A_{+}^{\dagger }\right) ^n}{%
\sqrt{n!}}D(-\beta )\left| 0\right\rangle \\
\left| n\right\rangle _{A_{-}} &=&\frac{\left( A_{-}^{\dagger }\right) ^n}{%
\sqrt{n!}}D(\beta )\left| 0\right\rangle ,
\end{eqnarray*}
where $D(\beta )=\exp \left( \beta a^{\dagger }-\beta a\right) $ is the
unitary displacement operator, acting on the original vacuum state $\left|
0\right\rangle $ ($a|0\rangle =0$). The states $\{\left| n\right\rangle
_{A_{+}}\}_n$, respectively $\{\left| n\right\rangle _{A_{-}}\}_n$, form an
orthonormal basis of the bosonic Hilbert space $L^2(\mathbb{R})$ and are
called extended coherent states (ECS)~\cite{chenqh}. Note that the hermitian
conjugated operators $A_{\pm}=(A^\dag_{\pm})^\dag$ annihilate the respective
displaced vacua $|0\rangle_{A_\pm}=D(\mp\beta)|0\rangle$.

The Hamiltonian in terms of $A_{+}^{\dagger }$, $A_+$ reads
\begin{widetext}
\begin{equation}
H_{1}=\left(
\begin{array}{ll}
A_{+}^{\dagger }A_{+}+\left( \lambda _{+}/\beta -\beta \right)
A_{+}^{\dagger }-\lambda _{+} & \;\;\;\;\left( -\frac{1}{2}\Delta +\lambda
_{-}\right) -\frac{\lambda _{-}}{\beta }A_{+}^{\dagger } \\
\;\;\;\;\left( -\frac{1}{2}\Delta -\lambda _{-}\right) +\frac{\lambda _{-}}{%
\beta }A_{+}^{\dagger } & \;A_{+}^{\dagger }A_{+}-\left( \beta +\lambda
_{+}/\beta \right) A_{+}^{\dagger }-2\beta A_{+}+2\beta ^{2}+\lambda _{+}%
\end{array}%
\right) .  \label{H_G}
\end{equation}%
\end{widetext}
The wavefunction can be expressed as the following series expansion using
these ECS
\begin{equation}
\left\vert A_{+}\right\rangle =\left(
\begin{array}{c}
\sum_{n=0}^{\infty }\sqrt{n!}e_{n}|n\rangle _{A_{+}} \\
\sum_{n=0}^{\infty }\sqrt{n!}f_{n}|n\rangle _{A_{+}}%
\end{array}%
\right) .  \label{wave_A+}
\end{equation}%
Multiplying the time-independent Schr\"odinger equation with the bra-vector $%
\;_{A_{+}}\left\langle m\right\vert \;$ from the left yields a recurrence
relation for the coefficients
\begin{widetext}
\begin{eqnarray}
e_{m} &=&\frac{\left( \beta -\frac{\lambda _{+}}{\beta }\right)
e_{m-1}+\left( \frac{1}{2}\Delta -\lambda _{-}\right) f_{m}+\frac{\lambda
_{-}}{\beta }f_{m-1}}{m-x},  \label{em} \\
f_{m} &=&\frac{\;\left( -\frac{1}{2}\Delta -\lambda _{-}\right) e_{m-1}+%
\frac{\lambda _{-}}{\beta }e_{m-2}+\left( m-1+2\beta ^{2}+2\lambda
_{+}-x\right) f_{m-1}-\left( \beta +\lambda _{+}/\beta \right) f_{m-2}}{%
2\beta m},  \label{fm}
\end{eqnarray}%
\end{widetext}
where $x=\lambda _{+}+E$ ($E$ is the energy). Starting from $f_{0}=1, f_m=0,
m<0$ we obtain all $f_{m}$ recursively.

The parity transformation $\hat{\Pi}$ maps the basis $\{\left|
n\right\rangle _{A_{+}}\}_n$ onto $\{\left| n\right\rangle _{A_{-}}\}_n$ and
vice versa, therefore the eigenfunction of $H_1$ in \eqref{wave_A+} can be
expressed also by the second type of ECS, if it is non-degenerate,
\begin{equation}
\left\vert A_{-}\right\rangle =\left(
\begin{array}{c}
\sum_{n=0}^{\infty }\left( -1\right) ^{n}\sqrt{n!}f_{n}|n\rangle _{A-} \\
\sum_{n=0}^{\infty }\left( -1\right) ^{n}\sqrt{n!}e_{n}|n\rangle _{A-}%
\end{array}%
\right) .  \label{wave_A-}
\end{equation}%

Both representations, \eqref{wave_A+} and \eqref{wave_A-} may differ only by
a multiplicative constant $z$
\begin{eqnarray}
\sum_{n=0}^{\infty }\sqrt{n!}e_{n}|n\rangle _{A_+} &=&z\sum_{n=0}^{\infty
}\left( -1\right) ^{n}\sqrt{n!}f_{n}|n\rangle _{A_-},  \notag \\
\;\sum_{n=0}^{\infty }\sqrt{n!}f_{n}|n\rangle _{A_+} &=&z\sum_{n=0}^{\infty
}\left( -1\right) ^{n}\sqrt{n!}e_{n}|n\rangle _{A_-}.  \label{prop}
\end{eqnarray}%
Multiplying these equations from the left with the original vacuum state ${%
\langle }0|$ and eliminating $z$ gives
\begin{equation}
\sum_{n=0}^{\infty }e_{n}\beta ^{n}\sum_{n=0}^{\infty }e_{n}\beta
^{n}=\sum_{n=0}^{\infty }f_{n}\beta ^{n}\sum_{n=0}^{\infty }f_{n}\beta ^{n},
\label{quad}
\end{equation}
where we have used
\begin{equation}
\sqrt{n!}{\langle }0|n{\rangle }_{A+}=(-1)^{n}\sqrt{n!}{\langle }0|n{\rangle
}_{A-}=e^{-\beta ^{2}/2}\beta ^{n},
\end{equation}%
Eq. \eqref{quad} can be written as
\begin{equation*}
G_+(x)G_-(x)=0
\end{equation*}
where we have defined the $G$-functions
\begin{equation}
G_{\pm }(x)=\sum_{n=0}^{\infty }\left( f_{n}\pm e_{n}\right) \beta ^{n}.
\label{G-function}
\end{equation}%
$E=x-\lambda_+$ is an eigenvalue of $H$ with positive (negative) parity if $%
G_+(x)$ ($G_-(x)$) vanishes at $x$. These $G$-functions are equivalent to
the $G$-functions obtained in Ref.~\cite{Fanheng} using the Bargmann space
in the sense that they have the same zeros as function of $E$.

\begin{figure}[tbp]
\center
\includegraphics[width=\linewidth]{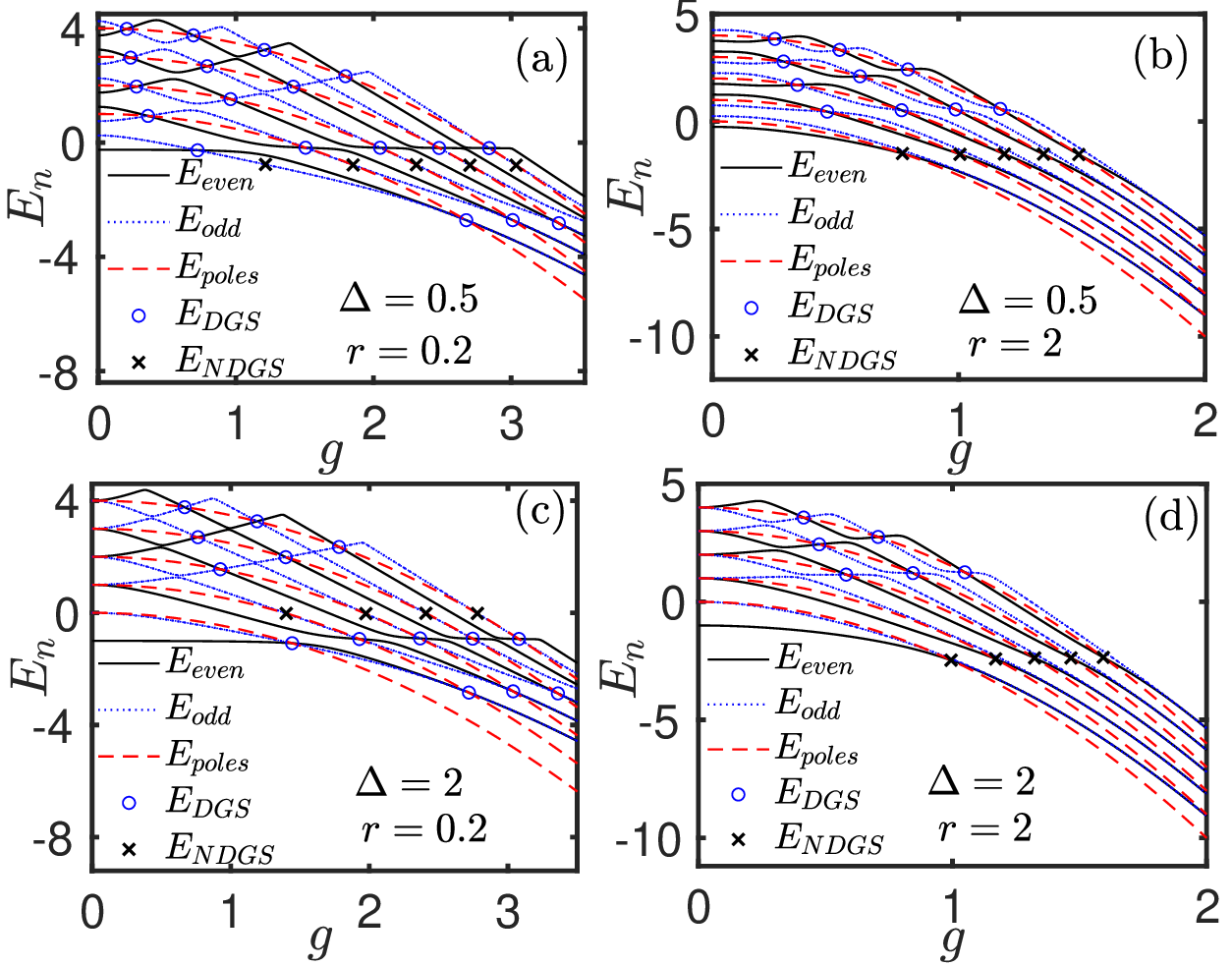}
\caption{ (Color online) The spectra and the isolated exceptional solutions
for the anisotropic QRM for $\Delta =0.5$ (upper panel) and $\Delta =2$
(lower panel), $r=0.2$ (left panel) and $r=2$ (right panel). Open circles
are obtained from Eq. (\protect\ref{excep}), crosses are determined by zeros
of the special G-functions Eq.~\eqref{G_non}. }
\label{energylevel}
\end{figure}

Let's emphasize that all non-degenerate eigenvalues (forming the regular
spectrum of the AiQRM) are given by the zeros of one of the $G$-functions in
Eq.~(\ref{G-function}), while for the (at most doubly) degenerate states Eq.~%
\eqref{quad} is not valid because Eq.~(\ref{prop}) presumes non-degeneracy,
i.e. the eigenstate must have a fixed parity. We will particularly pay
attention to the latter case in the next section, discussing the exceptional
spectrum. The energy spectra for $\Delta =0.5,2$ and $r=0.2,2$ obtained in
this way are presented in Fig.~\ref{energylevel}.

We will shown in the next section that the open circles correspond to
degenerate exceptional solutions and the crosses to non-degenerate
exceptional solutions. Both are accompanied by the lifting of a pole in
 both $G_+(x)$ and $G_-(x)$ (circles) or in only one of them
(crosses). The latter solutions can be obtained by a transcendental $G$%
-function, given in Eq.~\eqref{G_non}, dedicated to the non-degenerate
exceptional spectrum~\cite{Wakayama,braak-fmi} while the former are
characterized by algebraic equations for the model parameters (the
quasi-exact or ``Juddian'' solutions).

If $g_1=g_2=g$, the $G$-function of the isotropic QRM~\cite{Braak} is
recovered. Note that the ground state in the isotropic QRM  always has even
parity because it is continuously connected to the trivial case $g=0$ which
has even parity according to the definition of $\hat{\Pi}$.

\section{The exceptional spectrum}

From Eq. (\ref{em}), one notes that the coefficient $e_{n}$ diverges at $x=n$
due to its denominator if
\begin{equation}
E=n-\lambda _{+}.  \label{ex_e}
\end{equation}
Energies of this form are excluded from the regular spectrum, because they
corresponds to poles, not zeros of $G_\pm(x)$.
But let's assume there is nevertheless a state with energy $E=n-\lambda_+$.
In this case the numerator of the right-hand-side of Eq. (\ref{em}) should
vanish at $x=n$ so that $e_{n}$ remains finite, which results in two cases,
i.e. either (A),
\begin{equation}
\left( \beta -\frac{\lambda _{+}}{\beta }\right) e_{n-1}+\left( \frac{1}{2}%
\Delta -\lambda _{-}\right) f_{n}+\frac{\lambda _{-}}{\beta }f_{n-1}=0,
\label{excep}
\end{equation}%
with nonzero $e_{n-1},f_{n},$ and $f_{n-1}$, or (B),
\begin{equation}
e_{n-1}=0,f_{n}=0,f_{n-1}=0.  \label{non_deng}
\end{equation}%
These two requirements correspond either to a doubly degenerate eigenvalue
(case (A)) or a special non-degenerate state (case (B)). Both have the
energy $E=n-\lambda_+$ and are associated with the $n$-th pole line of $%
G_\pm(x)$. Energies of this form comprise the degenerate and non-degenerate
exceptional spectrum.

\subsection{Degenerate exceptional states}

\label{subA}

Equation (\ref{excep}) provides a constraint on the model parameters.
This constraint is actually a necessary and sufficient condition for the
occurrence of a doubly degenerate eigenvalue without specified parity,
because the pole at $x=n$ is lifted in both $G_+(n)$ and $G_-(n)$ and
neither function vanishes or diverges at the exceptional energy parameter $%
x=n$. The model parameters $\{g,\Delta,r\}$ satisfying Eq.~\eqref{excep}
indicate therefore a level crossing in the spectral graph located on the
pole line $E(g,r)=n-\lambda_+(g,r)$.

By eliminating the $e_i$, Eq. (\ref{excep}) can be written explicitly as
\begin{equation}
F_{n}(g)=\sum_{i=0}^{n}\Gamma _{i}\left( n\right)f_{i}=0  \label{condition}
\end{equation}%
where
\begin{equation*}
\Gamma _{i}\left( n\right) =\frac{\left( \frac{\lambda _{+}}{\beta }-\beta
\right) ^{\left( n-i\right) }}{\left( n-i\right) !}\left[ \left( \frac{n-i}{%
\lambda _{+}-\beta ^{2}}-1\right) \lambda _{-}+\frac{1}{2}\Delta \right]
\end{equation*}%
%
Given the parameters $\Delta $ and $r$, the coupling strength $g_k^{(n)}$
where the doubly degenerate state on the $n$-the pole line occurs follows
from Eq.~(\ref{condition}). In general, there is more than one solution for
given $n$, these solutions are labeled here with the index $k$. These states
are marked with open circles in Fig.~\ref{energylevel}. Let us analyze these
doubly degenerate states in detail below.

For $n=0$, $x=0$, the solution is unique and given by
\begin{equation}
g_1^{(0)}=\sqrt{\frac{\Delta }{1-r^{2}}},  \label{ex_gc}
\end{equation}%
which is the same as Eq. (11) in Ref. \cite{Fanheng} using the Bargmann
space approach and Eq. (15) in Ref. \cite{Duan2015}. It follows that the
first excited state and the ground state can only intersect if the CRW
coupling is different and weaker than the RW one. The parity in the lowest
energy state switches passing through this point, so a first-order quantum
phase transition occurs at $g_1^{(0)}$, in sharp contrast with the isotropic
QRM. From Eq.~(\ref{ex_gc}), the first level crossing in the left panel of
Fig. \ref{energylevel} occurs at $g_{1}^{(0)}=0.7217$ and $1.4434$ for $%
r=0.2,$ {$\Delta =0.5$} and $2$, respectively, consistent with numerical
calculations. As shown in the right panel of Fig. \ref{energylevel}, the
first two levels do not cross. No real solutions exist for $r>1$.

For the second pole line, $n=1,x=1$, Eq.~(\ref{condition}) reduces to a
cubic equation for $y=g^{2}$ as
\begin{equation}
2y\left( 1+r^{2}\right) -1+\frac{\Delta ^{2}-y^{2}\left( 1-r^{2}\right) ^{2}%
}{4}+\frac{2}{\frac{\Delta }{y\left( 1-r^{2}\right) }-1}=0.  \label{pol1}
\end{equation}%
%
Unlike Eq. (\ref{ex_gc}), real solutions of Eq.~\eqref{pol1} and Eq. (\ref%
{condition}) for all $n>1$ exist for $r>1$.

\begin{figure}[tbp]
\center
\includegraphics[width=\linewidth]{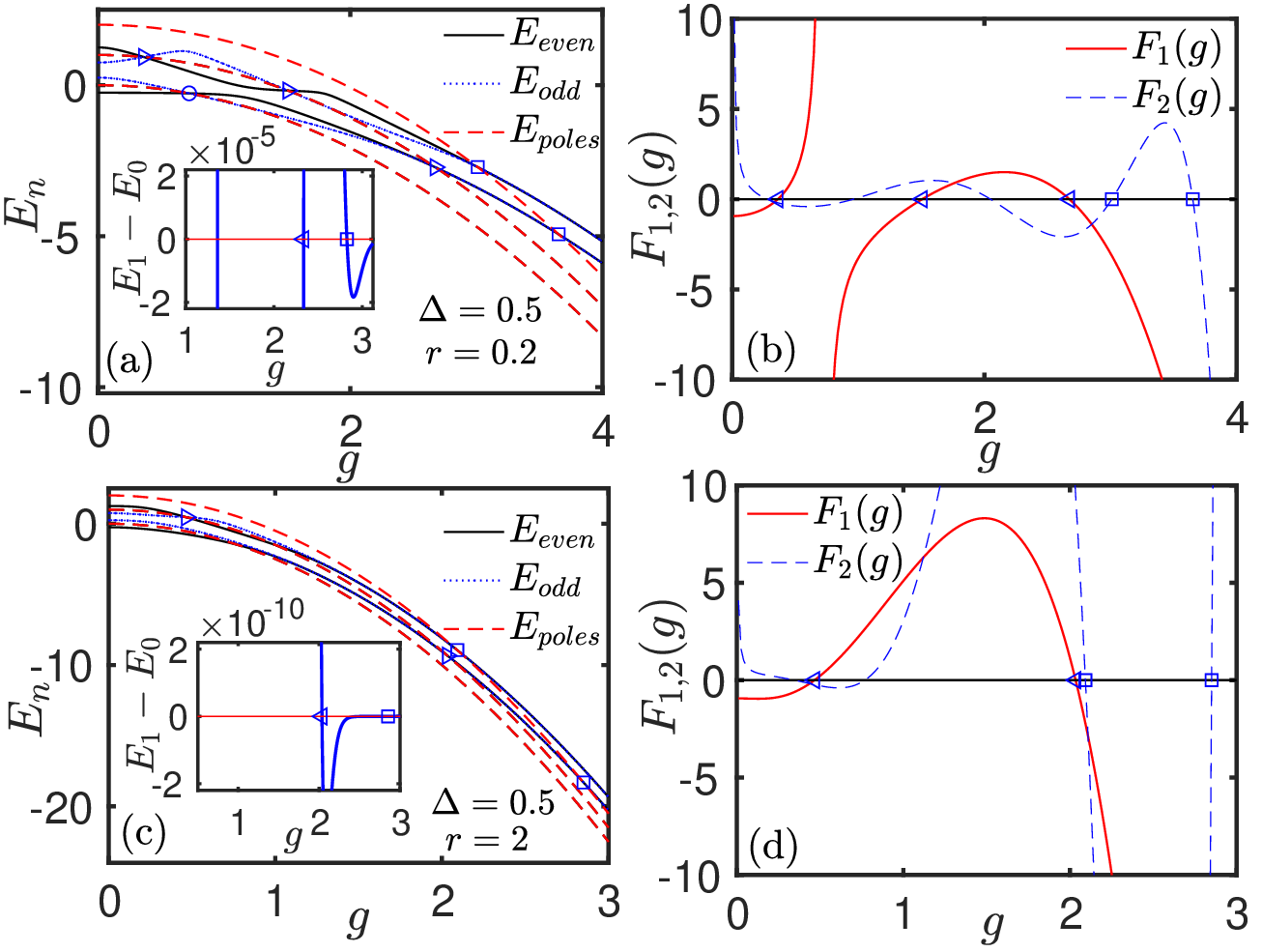}
\caption{ (Color online) The spectra for the first $4$ levels and the
degenerate exceptional solutions for the anisotropic QRM with anisotropic
constant $r=0.2$ (upper panel) and $r =2$ (low panel) for {  $\Delta =0.5$.}
Solutions to Eq. (\protect\ref{condition}) are marked with open symbols: $%
n=0 $ (circle), $n=1$ (triangle), and $n=2$ (square). The zeros of $%
F_{1,2}(g)$ corresponding to open symbols in the left panels are indicated
by the same symbols in the right panels. Note that some zeros associated
with level crossings of states above those present in the left panels are
not marked with symbols. }
\label{EnD2r01}
\end{figure}

Surprisingly, as shown in the left panel of Fig.~\ref{energylevel}, the
first two levels seem to intersect the pole line $n=1$ after they have
intersected the line $n=0$ at $g_1^{(0)}$. Is this a true level crossing or
a numerical artefact? To this end, we replot the spectral graph for {  $\Delta =0.5$},  $r=0.2$ and $2$, for the first four levels in Figs.~\ref{EnD2r01}~(a) and ~\ref{EnD2r01}~(c), respectively.
In Figs.~\ref{EnD2r01}~(b) and (d) we plot the functions $F_1(g)$ and $F_2(g)
$ whose zeros give the position of the couplings where the level crossings
occur. This shows that although the crossings are barely recognizable in the
plots of the left panels, only revealed on an enlarged scale (insets), the
functions in the right panels exhibit the double degenerate points clearly.
On sees that for $r<1$ the ground state parity changes at level crossings
with the first excited state at the pole lines $n=0,1,2$ for increasing $g$,
whereas the parity change occurs also for $r>1$ but at level crossings
belonging to $n=1,2$, as the line $n=0$ does not support a degeneracy for $%
r>1$. These quantum phase transitions for $r>1$ happen only in the deep
strong coupling limit (for $r=2$, $g/g_c\sim 6$), where the ground state and
the first excited state are almost degenerate anyway.

\begin{figure}[tbp]
\center
\includegraphics[width=\linewidth]{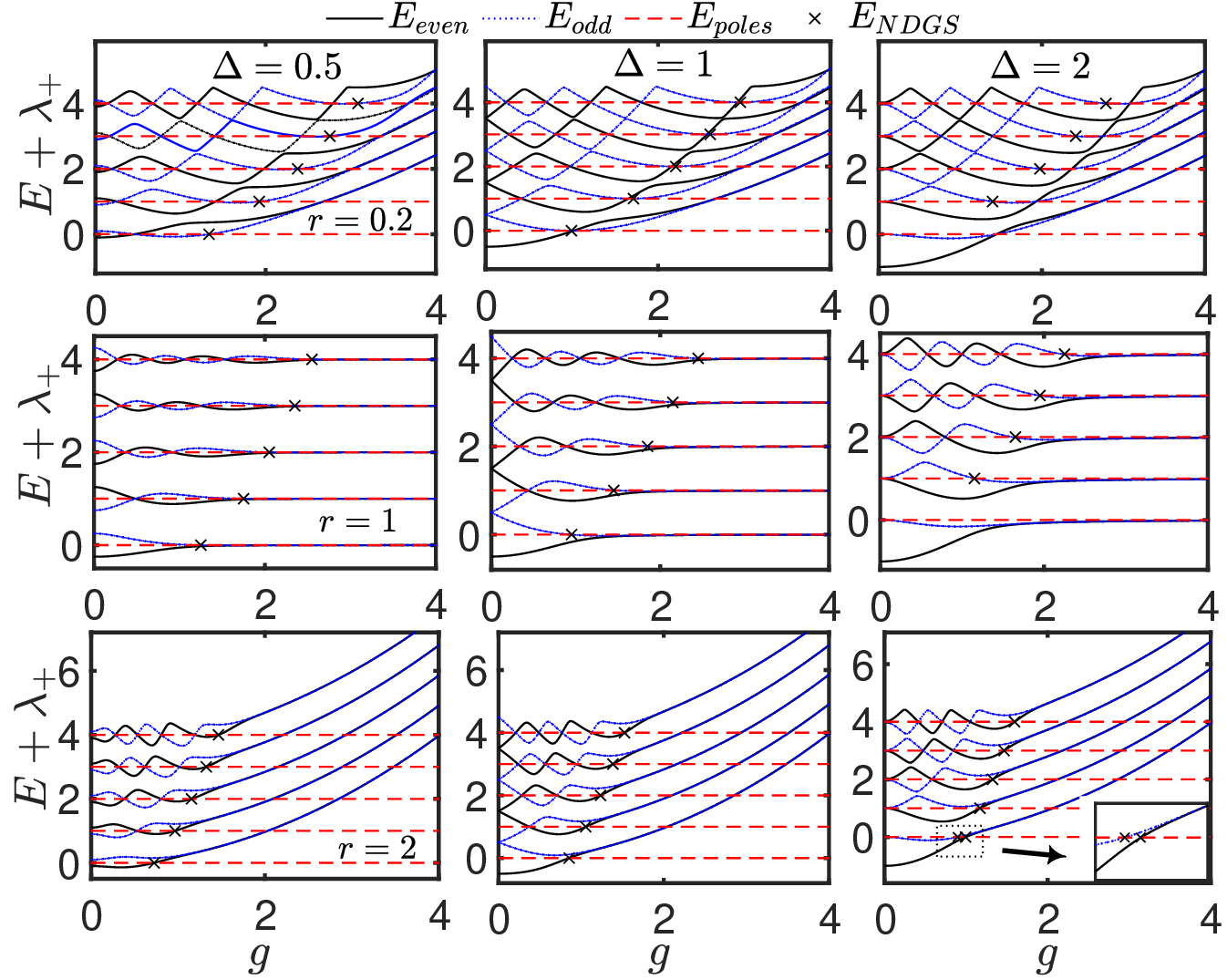}
\caption{ (Color online) The spectra $E+\protect\lambda _{+}$ for the
anisotropic QRM at $\Delta =0.5$ (left panel), $1$ (middle
panel), and $2$ (right panel). $r=0.2$ (upper panel), $r=1$ (middle panel),
and $r=2$ (lower panel). Crosses denote the non-degenerate solutions
determined by zeros of the special G-functions Eq.~\eqref{G_non}. }
\label{spectra}
\end{figure}

To show the level crossings more clearly, we present the energy spectra $%
E+\lambda _{+}$ as a function of the coupling $g$ in Fig.~\ref{spectra}. The
pole lines are now horizontal(red dashed lines). The most interesting
feature of the AiQRM, which sets it apart from the isotropic QRM, is the
fact that for increasing $g$, the ground state rises in energy compared to
the pole lines. In the isotropic case $r=1$, the ground state never crosses
the first pole line $n=0$ (middle panels in Fig.~\ref{spectra}) and the
asymptotic energies in the deep strong coupling regime are the pole energies
(all energies are asymptotically doubly degenerate). In the AiQRM for $r\neq
1$, the GS energy crosses all pole lines for $n>1$ if the coupling grows and
this allows for a succession of first order phase transitions if these
crossings belong to the degenerate exceptional spectrum. This is not
necessary the case, because they could also be non-degenerate exceptional
solutions, discussed in Sec.~\ref{subB}. However, we find that the
largest zero $g^{(n)}_{\text{max}}$ of $F_n(g)$ always corresponds to a
crossing of the ground state and the first excited state because all
non-degenerate exceptional solutions $g^{(n)}_{\text{n.d.}}$ for given $n$
satisfy $g^{(n)}_{\text{n.d.}} < g^{(n)}_{\text{max}}$.

\textsl{Quasi-exactness of the doubly degenerate eigenvalues:-}  For a doubly
degenerate eigenvalue $E=m-\lambda _{+}$, the model parameters satisfy Eq. (%
\ref{excep}). This means that $e_{m}$ is not determined by the recurrence relations and
may take an arbitrary value before normalizing the wave function. If we
fix it to
\begin{equation*}
\;e_{m}=\frac{\left[ \frac{\lambda _{-}}{\beta }e_{m-1}+\left( 2\beta
^{2}+2\lambda _{+}\right) f_{m}-\left( \beta +\lambda _{+}/\beta \right)
f_{m-1}\right] }{\left( \frac{1}{2}\Delta +\lambda _{-}\right) },
\end{equation*}%
we have $f_{m+1}=e_{m+1}=0$. It can be easily shown that all coefficients $%
f_{k}$ and $e_{k}$ for $k>m$ vanish whereas the coefficients $e_{n},f_{n}$
for $n<m$ are defined by the recurrence relations \eqref{em} and \eqref{fm}.
This leads to the expression of one of the doubly degenerate eigenfunctions
in terms of the first ECS,
\begin{equation}
\left\vert A_{+}\right\rangle _{m}=\left(
\begin{array}{c}
\sum_{n=0}^{m}\sqrt{n!}e_{n}|n\rangle _{A_{+}} \\
\sum_{n=0}^{m}\sqrt{n!}f_{n}|n\rangle _{A_{+}}%
\end{array}%
\right) .  \label{wave_PAM}
\end{equation}%
This eigenfunction has no fixed parity, but it can be written as a finite
polynomial in the shifted oscillator states $|n\rangle _{A_{+}}$. Applying
now the parity operator $\hat{\Pi}$ to this state, we obtain
\begin{equation}
\left\vert A_{-}\right\rangle _{m}=\left(
\begin{array}{c}
\sum_{n=0}^{m}\left( -1\right) ^{n}\sqrt{n!}f_{n}|n\rangle _{A-} \\
\sum_{n=0}^{m}\left( -1\right) ^{n}\sqrt{n!}e_{n}|n\rangle _{A-}%
\end{array}%
\right) ,  \label{wave_NAM}
\end{equation}%
which is obviously another state with energy $m-\lambda _{+}$, linearly
independent from $\left\vert A_{+}\right\rangle _{m}$. Both states have a
finite expansion in their respective ECS bases but the expansion in the
original bosonic Fock states does not terminate, in contrast to the
\textquotedblleft dark states\textquotedblright\ occurring in the Dicke
models \cite{jie}.

In the isotropic QRM, the proof is even simpler. If $E=m-g^{2}$, which is
the $m$-th pole energy, then the condition for double degeneracy reads
\begin{equation}
f_{m}(x)=0,
\end{equation}%
using
\begin{equation*}
\left( m-g^{2}-E\right) e_{m}=\frac{\Delta }{2}f_{m},
\end{equation*}%
Thus $e_{m}$ would be arbitrary. If we set
\begin{equation*}
e_{m}=-\frac{4g}{\Delta }f_{m-1},
\end{equation*}%
then $f_{m+1}=0$, further $e_{m+1}=0$. Because both $f_{m}$ and $f_{m+1}$
are zero, $f_{m+2}$ and $e_{m+2}$ vanish as well. In this case the
coefficients $f_{k}$ and $e_{k}$ for $k>m+1$ vanish, thus one of the
degenerate eigenfunctions is given as a finite polynomial in the ECS basis $%
\{\left\vert n\right\rangle _{A_+}\}$ (see also~\cite{Braak19}). These
states are the quasi-exact solutions of the QRM found originally by Judd~%
\cite{Judd,Koc}.

\begin{figure}[tbp]
\center
\includegraphics[scale=0.45]{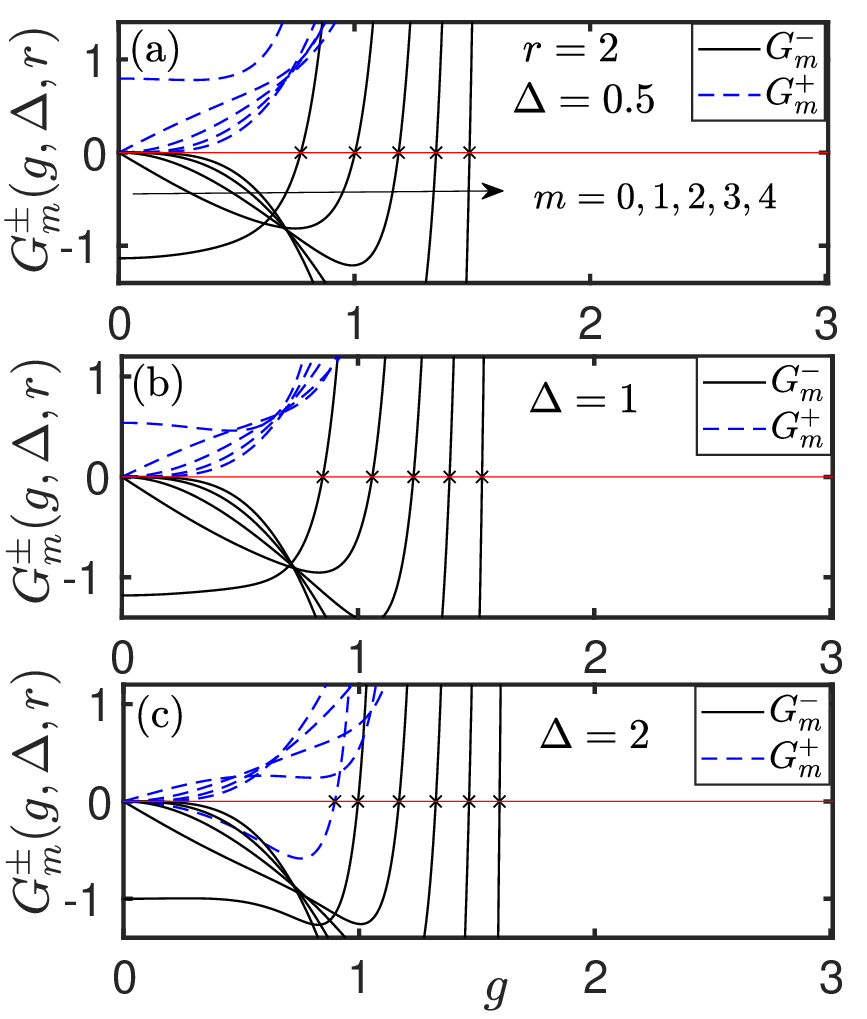}
\caption{ (Color online) The special $G$-functions $G_m^\pm(g,\Delta,r)$ for
$r=2$. Their zeros give the non-degenerate exceptional solutions $g^{(m)}_{\text{n.d.}}$, i.e. crossings with the pole line $m$. These are depicted as
crosses in the lower panel of Fig.~\protect\ref{spectra}. Note that for
large $\Delta$, the line $m=0$ supports non-degenerate exceptional solutions
of both parities for $g\sim 1$.}
\label{nondegs}
\end{figure}

\subsection{Non-degenerate exceptional states}

\label{subB}

The non-degenerate exceptional states correspond also to states with $%
E=m-\lambda_+$ but this energy is non-degenerate because the pole at $x=m$
is only lifted in one of the $G$-functions, $G_+(x)$ or $G_-(x)$, but not in
both. The state is therefore a parity eigenstate. These states have been
first analyzed for the QRM with the Bargmann space technique in \cite%
{Maciejewski21} and later in \cite{Braak19} and \cite{braak-fmi}.

We shall analyze them now with the ECS technique for the AiQRM. If condition
(\ref{non_deng}) holds, $e_m$ can take an arbitrary value, while all
coefficients $e_k,f_k$ for $k<m$ vanish.
Setting $e_{m}=1$, $f_m=0$ all coefficients for $e_{n}$ and $f_{n}$ with $%
n\ge m+1$ are fixed by the recurrence relations (\ref{em}) and (\ref{fm}).
Imposing now that the constructed state is a parity eigenstate, we find that
one of the $G$-functions
\begin{equation}
G_{m}^{\pm}(g,\Delta,r)=\pm \beta ^{m}+\sum_{n=m+1}^{\infty }\left( f_{n}\pm
e_{n}\right) \beta ^{n}  \label{G_non}
\end{equation}%
must vanish. These $G$-functions are associated with the exceptional energy $%
m-\lambda_+$ and the $m$-th pole line. They are functions of the model
parameters and their vanishing puts a constraint on these parameters,
similarly to Eq.~\eqref{condition} for the degenerate eigenvalues.
The non-degenerate exceptional eigenstates are marked by crosses in Fig. \ref%
{energylevel} and correspond to zeros of either $G_m^+(g,\Delta,r)$ or $%
G_m^-(g,\Delta,r)$. They are not quasi-exact states like the
doubly-degenerate eigenstates, because the functions $G^\pm_m(g,\Delta,r)$
are not polynomials in $\beta$.

 The $G$-functions $G_m^\pm(g,\Delta,r)$, $m=0,\ldots 4$, for several $\Delta$ at $r=2$ are plotted in Fig.~%
\ref{nondegs}  (c.f. the lower panel in Fig.~\ref{spectra}). We see that each of them has at most
one zero at the coupling $g^{(m)}_{\text{n.d.}}$, which happens to be always
smaller than $g^{(m)}_{\text{max}}$. While it would be interesting to show
this conjecture analytically, we confine ourselves here to a numerical check. It
entails that the AiQRM exhibits an infinite series of phase transitions for
increasing coupling, similar to the Jaynes-Cummings model, not only for $r<1$%
, where the RW terms dominate but also in the dual case $r>1$.

\section{Ground state instability}

\begin{figure}[tbp]
\center
\includegraphics[width=\linewidth]{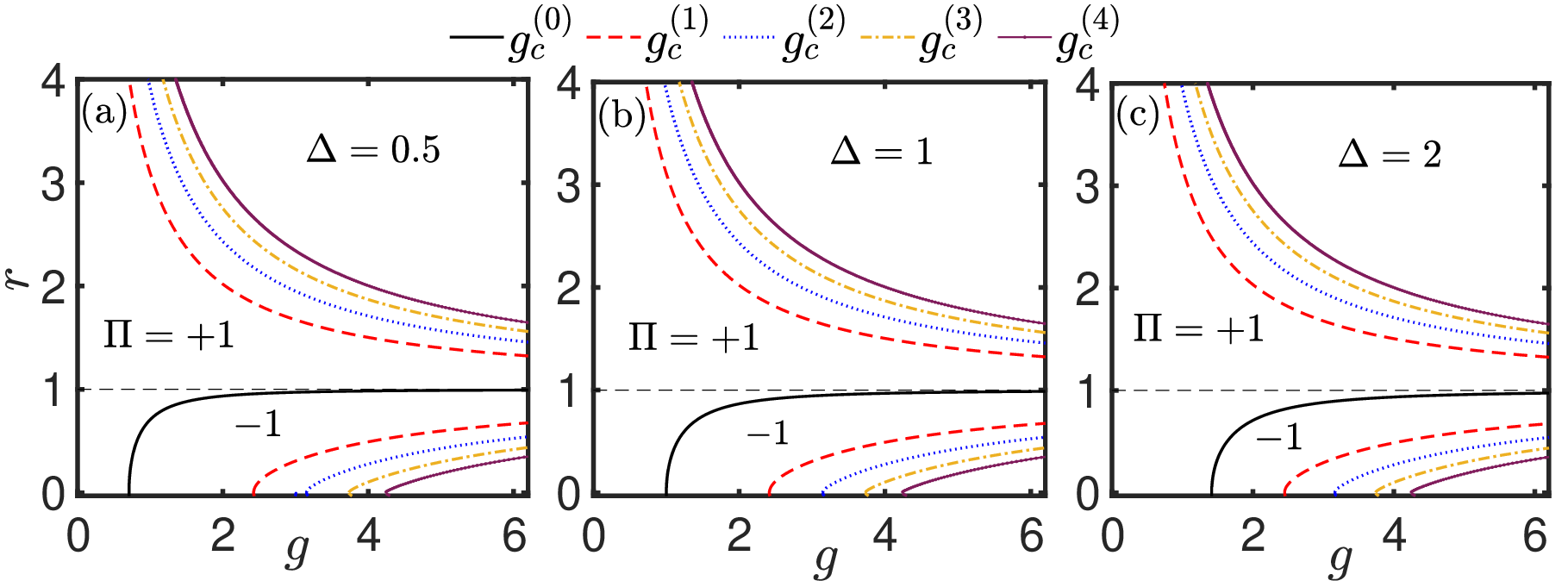}
\caption{ (Color online) The phase diagram in the $r/g$ plane for the AiQRM
at $\Delta  =0.5$ (left panel), $1$ (middle panel), and $2$
(right panel). $g_c^{(n)}$ is the maximum among the solutions $g_k^{(n)}$ of
Eq.~(\protect\ref{condition}) for given $n$. The isotropy line $r=1$
(dashed) separates the two regions of the phase diagram where multiple
quantum phase transitions are possible. The line $r=0$ denotes the
Jaynes-Cummings model.}
\label{phase}
\end{figure}

The intersections of energy levels in the spectral graph are directly
related to the symmetries of the Hamiltonian. In the case of the isotropic
QRM and the AiQRM, we have seen that all level crossings in these models
have the same origin, namely the manifest $\mathbb{Z}_2$-symmetry. The crucial
feature of these degeneracies is that they are located always on the lines
where the $G$-functions have poles. If the ground state energy crosses one
of these lines, as is the case for any non-vanishing anisotropy ($r\neq 1$),
the concomitant degeneracy of the ground state indicates a quantum phase
transition of first order, where the symmetry of the ground state is not
defined. The otherwise well-defined parity of the ground state changes sign
upon crossing these points. Because the ground state energy crosses
eventually \textit{all} pole lines in the AiQRM, the system undergoes
infinitely many such phase transitions for increasing coupling. The ground
state phase diagram in the $g/r$-plane is shown in Fig.~\ref{phase} for
three different values of $\Delta$. The parity of the ground state in
the different phases is either $+1$ or $-1$. The parity is unique and
positive in a region around the isotropy line $r=1$, while we recover the
infinitely many phase transitions of the Jaynes-Cummings model for $r=0$.
The phase diagram for $r<1$ is consistent with that proposed recently by
Ying using numerical exact diagonalization~\cite{yingzj}.

In principle, the crossing of an energy level with a pole line could be due
to a non-degenerate exceptional eigenstate [as in Figs.~\ref{energylevel}
~(b) and ~\ref{energylevel} (d)] and would not indicate a quantum phase transition. However, our
numerical checks have shown so far that all maximal solutions of Eq.~%
\eqref{condition} belong indeed to a degeneracy of the ground state,
although most of them occur in the deep strong coupling regime, where we
have already almost perfect degeneracy of the two lowest energy eigenstates.

Finally, we would like to point out that the switch between even and odd
parity of the ground state has been observed in the anisotropic spin-boson
model by two of the present authors \cite{yanzhi}. As shown in the top area
of Fig. 1(a) in that paper, a delocalized phase with even parity switches to
a delocalized phase with odd parity with increasing coupling strength in the
highly anisotropic system.

\section{Conclusions}

In this work, we derive the two parity $G$-functions for the anisotropic
quantum Rabi model employing its manifest $\mathbb{Z}_2$-symmetry by the
Bogoliubov operator (ECS) approach in the physical Hilbert space $L_2(%
\mathbb{R})$. Zeros of the $G$-functions yield the regular spectrum with
well-defined parity. The exceptional solutions are located at the pole lines
of these $G$-functions and comprise all doubly degenerate eigenvalues. The
condition for their occurrence is derived in closed form.
This allows us to identify an infinity of first-order quantum phase
transitions in the AiQRM, whenever the model is not fully isotropic. The
importance of the analytical treatment becomes clear as in many cases the
numerical resolution of the spectra is very difficult, especially in the
deep strong coupling regime (see Fig.~\ref{energylevel}).

At each crossing of the two lowest energy states the parity of the ground
state switches between the discrete values $+1$ and $-1$ for increasing
coupling strength. For the extreme case $r=0$ (Jaynes-Cummings model), the
value of the excitation number $\hat{N}$ rises by $+1$ at each phase
transition point. While $\hat{N}$ is not conserved for $r\neq 0$, the infinite
number of phase transitions remains in the anisotropic case at any value $%
r\neq1$. The only model with no phase transition in the ground state for any
coupling is the isotropic QRM. The rich phase diagram of the AiQRM is solely
due to its manifest $\mathbb{Z}_2$-symmetry. In contrast, the level
crossings of higher excited states occurring at special values of the
symmetry-breaking parameter $\epsilon$ in the asymmetric QRM (where the $%
\mathbb{Z}_2$-symmetry is broken by the term $\epsilon\sigma_x$ in the
Hamiltonian) is due to a non-manifest, hidden symmetry \cite%
{bat,ash2020,man,rey}.

\begin{acknowledgments}
This work is supported by the National Science Foundation
of China under Grant No. 11834005, the National Key Research
and Development Program of China under Grant No.
2017YFA0303002, and by the German Research Foundation
(DFG) under Grant No. 439943572.
\end{acknowledgments}

$^{\dagger}$  daniel.braak@physik.uni-augsburg.de

$^{*}$ qhchen@zju.edu.cn

\end{document}